# Personalization of Basic and Moral Values Guiding LLM-Controlled Assistant Robots

Mateusz Woźniak[1,2[0000-0002-1693-7020]], Qi Gao[1[0000-0002-4984-877X]], and Agnieszka Wykowska[1[0000-0003-3323-7357]]

[1] Italian Institute of Technology, Via Enrico Melen 83, 16152 Genoa, Italy
[2] Institute of Applied Psychology, Jagiellonian University, ul. Łojasiewicza 4, 30-348 Cracow, Poland
mgwozniak@gmail.com, qi.gao@iit.it, agnieszka.wykowska@iit.it

**Abstract.** Personal assistant robots powered by large language models (LLMs) are likely to enter everyday life within the next decade, but the values that should guide their behaviour remain largely unspecified. Current alignment practice steers a single model toward an aggregate of human values, ignoring that individual users may prefer a robot whose value profile differs from their own. We administer parallel self- and robot-targeted versions of three established psychometric instruments – the simplified Schwartz Values Survey (SVS), the Portrait Values Questionnaire (PVQ-21), and the Moral Foundations Questionnaire (MFQ) – to potential future users (N=22), capturing both their personal values and the values they would want an assistant robot to embody. We compare the three instruments as personalization tools using the UEQ-S and benchmark fifteen contemporary LLMs against the resulting profiles. Participants wanted their robot to score markedly lower than themselves on personal-focus values and higher on social-focus values; moral profiles were nearly identical. LLMs prompted to act as robots shifted only weakly toward the desired profile, indicating that default model behaviour and prompt engineering are no substitute for explicit value personalization.

## 1 Introduction

Personal assistant robots are no longer a distant prospect. Advances in large language models (LLMs), embodied perception, and low-cost humanoid platforms make it plausible that within a decade many households will share their space with a robotic agent that schedules appointments, supports household management and activities, and engages in extended social interaction. Importantly, such a robot cannot be morally or motivationally neutral: every action and decision prioritizes some considerations over others: efficiency over comfort, autonomy over safety, the household's privacy over a guest's convenience. The decisions made in such trade-offs are motivated by what psychology has long studied under the terms of basic values [16] and moral values [9]. If an assistant robot is to autonomously act in ways its user finds acceptable, its priorities must align with the user's in a way that reflects such values.

Current alignment practice addresses this challenge at the population level. Reinforcement learning from human feedback and instruction tuning steer a single model toward an aggregate of human preferences, producing one value profile that ships to every user. Recent AI-psychometric work [5, 11, 13] confirms that this average profile is non-neutral: contemporary LLMs systematically over-prioritize universalism and self-direction and under-emphasize achievement, power, and security relative to humans, and individual models differ measurably from one another. An assistant robot, however, is used by an individual person rather than by a population, and the question of whether users might want a robot that is guided by a custom profile of values – which might reflect their own or not – remains largely open.

We directly address this personalization gap. We administer parallel self- and robot-targeted versions of three established psychometric instruments – the simplified version of the Schwartz Values Survey (SVS), the Portrait Values Questionnaire (PVQ, version PVQ-21), and the Moral Foundations Questionnaire (MFQ) – to a sample of potential future users, capturing in a single session both their own values and the values they would want for an assistant robot. We accompany each instrument with the short User Experience Questionnaire (UEQ-S) so that the three can be compared as personalization tools, and we administer the same instruments to fifteen contemporary LLMs, both unprompted and prompted to act as a personal assistant robot.

## 2 Related works

### 2.1 Basic and moral values

The dominant framework for measuring basic personal values is Schwartz's Theory of Basic Values [16, 18], which organizes ten motivationally distinct values—self-direction, stimulation, hedonism, achievement, power, security, conformity, tradition, benevolence, and universalism into four higher-order dimensions (openness-to-change vs. conservation; self-enhancement vs. self-transcendence), as well as two orientations (personal-focus and societal-focus). It is operationalized by the SVS [16] and the PVQ, later refined into the PVQ-RR [20]. Moral Foundations Theory [8, 9] complements Schwartz by partitioning moral values into five dimensions: Care/Harm, Fairness/Cheating, Loyalty/Betrayal, Authority/Subversion, and Sanctity/Purity, measured with the MFQ [3]. Both frameworks have demonstrated cross-cultural validity and are well suited to embedding in computational systems.

### 2.2 Values in large language models

A growing AI-psychometric literature has shown that these instruments can be administered to LLMs and yield reliable value profiles. Pellert et al. [13] argue that LLMs inadvertently acquire psychological traits from their training data, and demonstrate that these can be elicited as readable response profiles. Within Schwartz's framework, Hadar-Shoval and colleagues [4–6] show that four leading LLMs produce distinct, instrument-stable motivational profiles that diverge from published human norms – over-prioritizing universalism and self-direction, under-emphasizing achievement, power, and security – and that these profiles predict the models' choices on ethical dilemmas. Liu et al. [11] extend the picture to over twenty models from China- and West-origin providers, reporting systemic instability in LLM value systems and a non-trivial relationship between model scale and cross-cultural alignment. Smith-Vaniz et al. [21] obtain analogous results for values from Moral Foundations Theory. A complementary line of work treats values as a training signal: Qiu et al. [14] use VALUENET to make dialogue agents more consistent with a target value persona, and Watson et al. [22] propose a per-user fine-tuning framework. Both confirm that value-level personalization is technically feasible, but neither has been integrated with validated psychometric instruments or evaluated in embodied HRI.

### 2.3 Values in human–robot interaction

In human–robot interaction, the role of value has been investigated in regard to moral values. Čaić et al. [1] argue that users' personal values become salient through interactions with social robots and argue that a robot whose values are miscalibrated to the user will be evaluated more poorly regardless of objective performance. The European VALAWAI project further focuses on developing "value-aware AI" systems that can understand, align with, and act according to human norms and ethical frameworks. This project shows that humans attribute moral values to robots and evaluate their behavior based on perceived value alignment [12]and people weigh social norms more than other values when judging robot decision [2]. However, to our knowledge, no empirical study has measured users' own and desired-robot value profiles with the same validated instrument.

### 2.4 Research questions

Three gaps motivate the present study, which we address through three research questions:

**RQ1**. Do users want an assistant robot whose values mirror their own, complements them, or approximates a population average?

**RQ2.** How does the desired-robot value profile compare with the default value profile of contemporary LLMs, both unprompted and when prompted to act as a personal assistant robot?

**RQ3.** How do the SVS, PVQ, and MFQ compare head-to-head as instruments for eliciting an individual user's desired robot value profile?

# 3 Methods

## 3.1 Participants

The study involves administration of a series of questionnaires to a group of participants (N=22) mostly recruited at the Italian Institute of Technology (IIT) in Genoa, Italy.

Among 22 participants 13 identified as female and 9 as male. Their mean age was 30.7 years (*SD*=6.0, min=22, max=43).

All participants gave informed consent after reading a general information about the study and before starting the experiment. The study was approved by the local ethical committee (Comitato Etico Regione Liguria). The data were collected at the Italian Institute of Technology, IIT in Genoa, Italy, in accordance with the standards from the Declaration of Helsinki.

## 3.2 Procedure

The study involved administration of a series of questionnaires, which are described in more detail in the next section. Participants started the study by reading the consent form and agreeing to it. Afterwards they were asked to provide their age and gender. They were also asked the following two questions: "Does your work involve robots (HRI, Robotics)?" and "Does your work involve scientific research?" The available answers to both questions were "yes" and "no". Afterwards they read the following instructions: "Imagine that you want to buy a humanoid robot that will serve as your personal assistant. On the following pages your task will be to answer a series of questions about how the robot should behave in different situations and what rules should guide its behavior. When answering these questions think about a robot that you would like to buy and use."

Afterwards, participants completed three questionnaires in which they had to personalize the values of an assistant robot. The three questionnaires were: SVS-robot, PVQ-robot, and MFQ-robot. These questionnaires are described in more detail in section 3.2. Moreover, after completing each of them participants had to evaluate them by answering five custom questions, as well as completing the UEQ-S questionnaire.

As the next step they completed these same three questionnaires, but as applied to themselves, i.e. meaning that they had to evaluate to what extent these values are important for them personally. The order of presentation of the robot questionnaire block and questionnaire for themselves block was random. Within the two blocks, the order of the three questionnaires was also random.

## 3.3 Questionnaires

The study involved administration of the following questionnaires.

**Portrait Value Questionnaire (PVQ-21)** is the 21-item short form of the Portrait Values Questionnaire originally introduced by Schwartz et al. [19]. Each item is a third-person portrait of a person whose goals, aspirations or wishes implicitly express one of Schwartz's ten basic values (self-direction, stimulation, hedonism, achievement, power, security, conformity, tradition, benevolence, universalism); the respondent is asked, on a six-point Likert scale ("not like me at all" – "very much like me"), how similar they are to the described person. Scores on each value are computed as the mean of the corresponding two or three items [17, 20].

**Portrait Value Questionnaire (PVQ-robot)**: a robot-targeted adaptation in which the 21 portraits are rewritten in third-person neutral to describe an assistant robot (e.g., "It is important to it to…"), preserving the value content and Likert scale. Participants rate how similar each portrayed robot is to the assistant robot they would want, yielding a desired-robot profile on the same 10-value structure as the human PVQ-21.

**Schwartz Values Survey - simplified version of SVS:** an abbreviated, plain-language version of Schwartz's original Values Survey [16, 18], constructed for this study to keep the explanations easy to understand for general-population participants (see Table 1). Rather than rating fifty-six values from the standard SVS, respondents rated the ten basic Schwartz values, each presented with a one-sentence definition, on scale that runs from −5 ("extremely opposes my values") through 0 ("not important") to 5 ("extremely important"). Scores are taken as the raw rating per value [16].

**Table 1.** Items from the custom version of the SVS questionnaire, with definitions reformulated to make them easier to understand by users.

| Value | Definition as presented to participants |
|---|---|
| **Benevolence** | Being kind and caring about the well-being of people around you (your ingroup), trying to help them. |
| **Universalism** | Caring about all people (not just those you know) and about nature, and wanting fairness, understanding, and protection for everyone. |
| **Conformity** | Trying not to do things that upset others or break social rules. |
| **Tradition** | Valuing and respecting customs, beliefs, and practices passed down by culture or religion. |
| **Security** | Wanting safety, order, and stability in society, in relationships, and in one's own life. |
| **Power** | Wanting influence, authority, or control over people, resources, or situations. |
| **Achievement** | Wanting to succeed and show your abilities in ways that are recognized and valued by others. |
| **Hedonism** | Seeking pleasure, enjoyment, and physical satisfaction. |
| **Stimulation** | Wanting excitement, new experiences, and challenges. |
| **Self-Direction** | Valuing independence in thinking and acting, enjoying making your own choices, creating things and exploring new ideas. |

**Schwartz Values Survey - simplified version SVS-robot:** a parallel robot-targeted version using the same ten value definitions and metric scale; participants rate how important each value should be as a guiding principle for their assistant robot, yielding a desired-robot profile directly comparable with the human SVSs.

**Moral Foundations Questionnaire MFQ:** the leading instrument for measuring moral values, operationalizing Moral Foundations Theory [7–9]. It consists of two parts of sixteen items each (32 items total). Part 1 (Relevance) asks the respondent how relevant a given consideration (e.g., "whether or not someone suffered emotionally") is when judging whether something is right or wrong, on a 0–5 scale. Part 2 (Judgement) asks how strongly the respondent agrees with each of sixteen value-laden statements, on a 0–5 scale. Each part contains three scored items plus one validity-check filler item per foundation, yielding scores on the five moral foundations of Care/Harm, Fairness/Cheating, Loyalty/Betrayal, Authority/Subversion, and Sanctity/Purity [3].

**Moral Foundations Questionnaire MFQ-robot:** a robot-targeted adaptation retaining the two-part 32-item structure of the MFQ, with each item reframed to elicit the respondent's preferences for the assistant robot's moral reasoning. Scoring is unchanged, yielding a moral-foundations profile directly comparable with the participant's own.

**User Experience Questionnaire short UEQ-S:** The UEQ-S [15] is the eight-item form of the User Experience Questionnaire [10], designed for fast and reliable assessment of user-perceived quality. It comprises four pragmatic-quality items (obstructive–supportive, complicated–easy, inefficient–efficient, confusing–clear) and four hedonic-quality items (boring–exciting, not interesting–interesting, conventional–inventive, usual–leading-edge), each scored on a 7-point semantic-differential scale. Following [15], we report the pragmatic and hedonic sub-scale means and the overall UEQ-S score; in the present study the UEQ-S is administered separately for each of the value/morals questionnaires (PVQ, SVS and MFQ) so that the questionnaires themselves can be compared on perceived pragmatic and hedonic quality as personalization tools.

# 4 Results

## 4.1 Comparison of human and robot values

The first goal of the study was to investigate what value profile the users of assistant robots might request, and to what extent this profile differs from the value profile of the users themselves. To answer this question, we compared the own and desired value profile using three questionnaires. Repeated-measures t-test was conducted for each value. All p values are Bonferroni corrected unless stated otherwise.

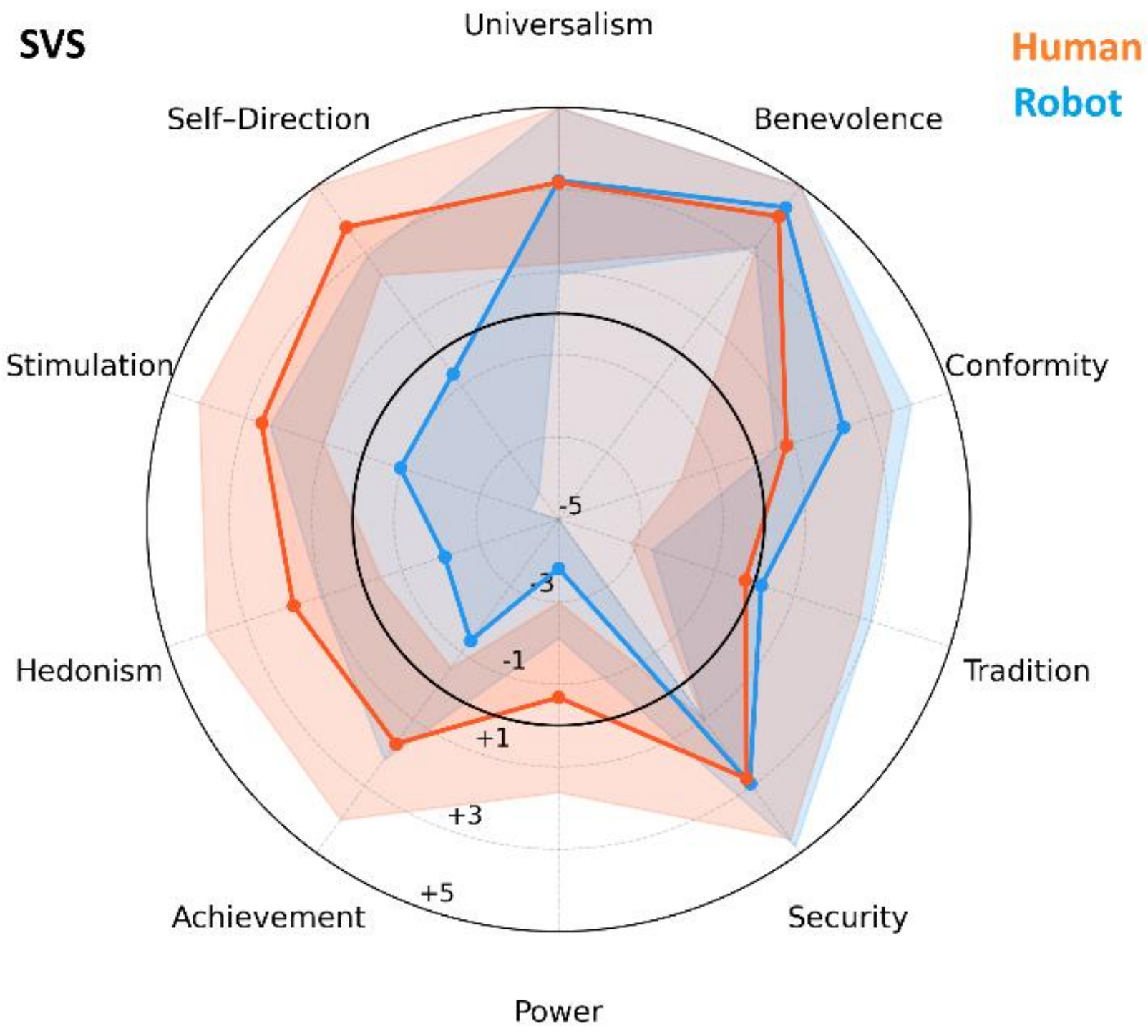

**Fig. 1.** Radar plot illustrating average reported own values (orange) and values that participants wanted their assistant robot to have (blue) as measured using the simplified version of the SVS.

**Schwartz Value Scale (simplified version)**. We found that when participants had to directly evaluate the importance of the values the value profile significantly differed between their own personal profile and the profile that they requested for a home assistant robot. Specifically, participants assigned much higher score for importance of power (t(21)=5.35, *p*<0.001, d=1.14), achievement (t(21)=4.46, *p*<0.001, d=0.95), hedonism (t(21)=5.95, *p*<0.001, d=1.27), stimulation (t(21)=5.66, *p*<0.001, d=1.21) and self-direction (t(21)=5.46, *p*<0.001, d=1.16) as values guiding their own behavior then the behavior of an assistant robot. At the same time they assigned higher importance to conformity (t(21)=2.81, *p*=0.010 uncorrected, d=0.60) for the robot. The other values did not significantly differ.

**Portrait Value Questionnaire (PVQ)**. In PVQ the differences between personal and robot values were less pronounced. Numerically the pattern of results was similar to SVS (lower scores for robot for values with personal focus, and higher for values with social focus), and we found similar significant differences for stimulation: (t(21)=4.18, *p*<0.001, d=0.89), self-direction: (t(21)=3.88, *p*=0.010, d=0.83), achievement (t(21)=4.18, *p*=0.050, d=0.67), and hedonism (t(21)=4.18, *p*=0.031 uncorrected, d=0.49) as we found in SVS.

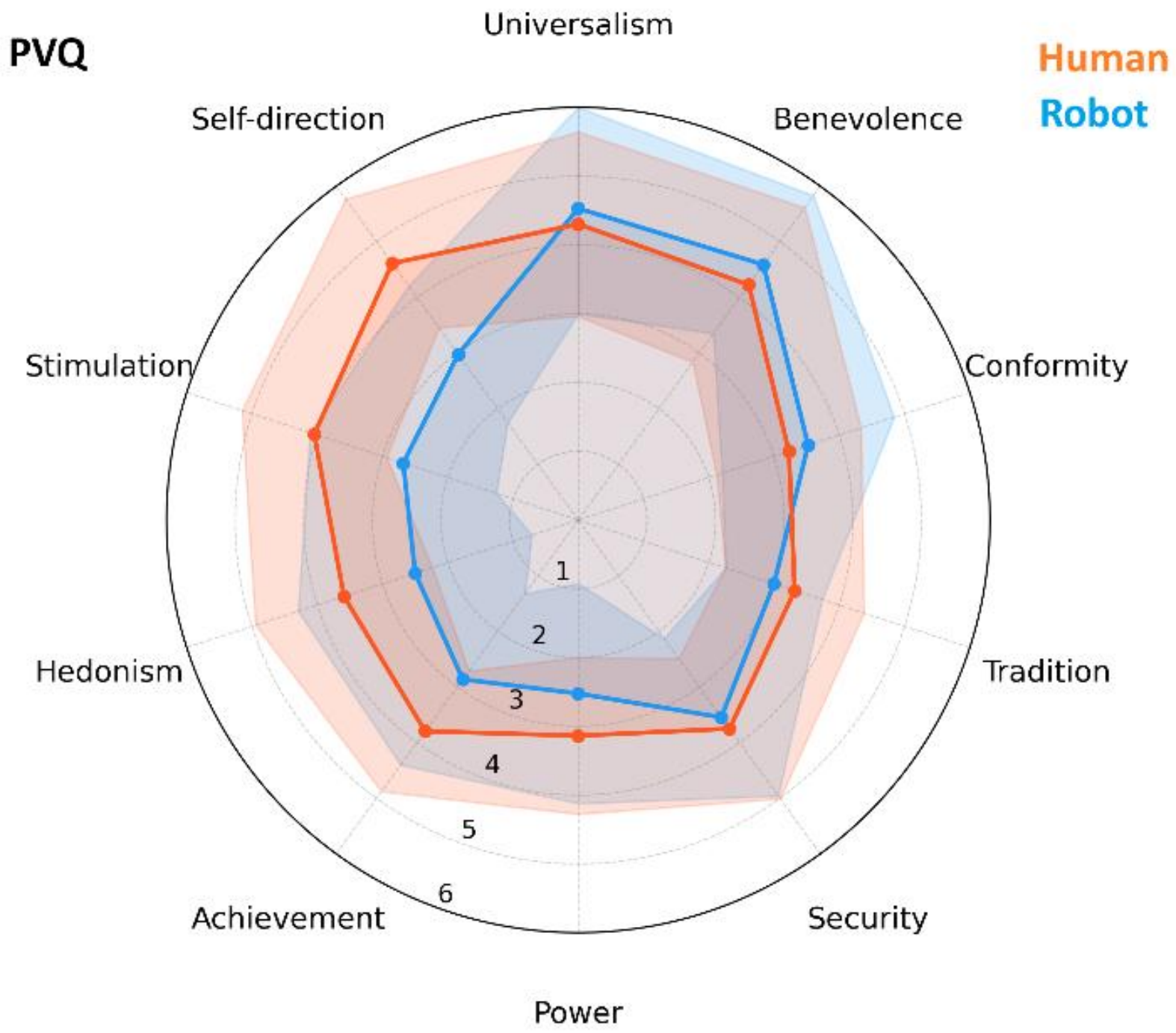

**Fig. 2.** Radar plot illustrating average reported own values (orange) and values that participants wanted their assistant robot to have (blue) as measured using the PVQ.

**Moral Foundations Questionnaire (MFQ)**. The MFQ investigates moral values rather than basic values and in contrast to basic values we observed much less differentiation between personal and robot values. Among the five moral dimensions we found a significant difference only for loyalty ($t(21)= 3.77$, $p=0.001$, $d=0.8$). Surprisingly, the importance of loyalty was rated as lower for robot than human.

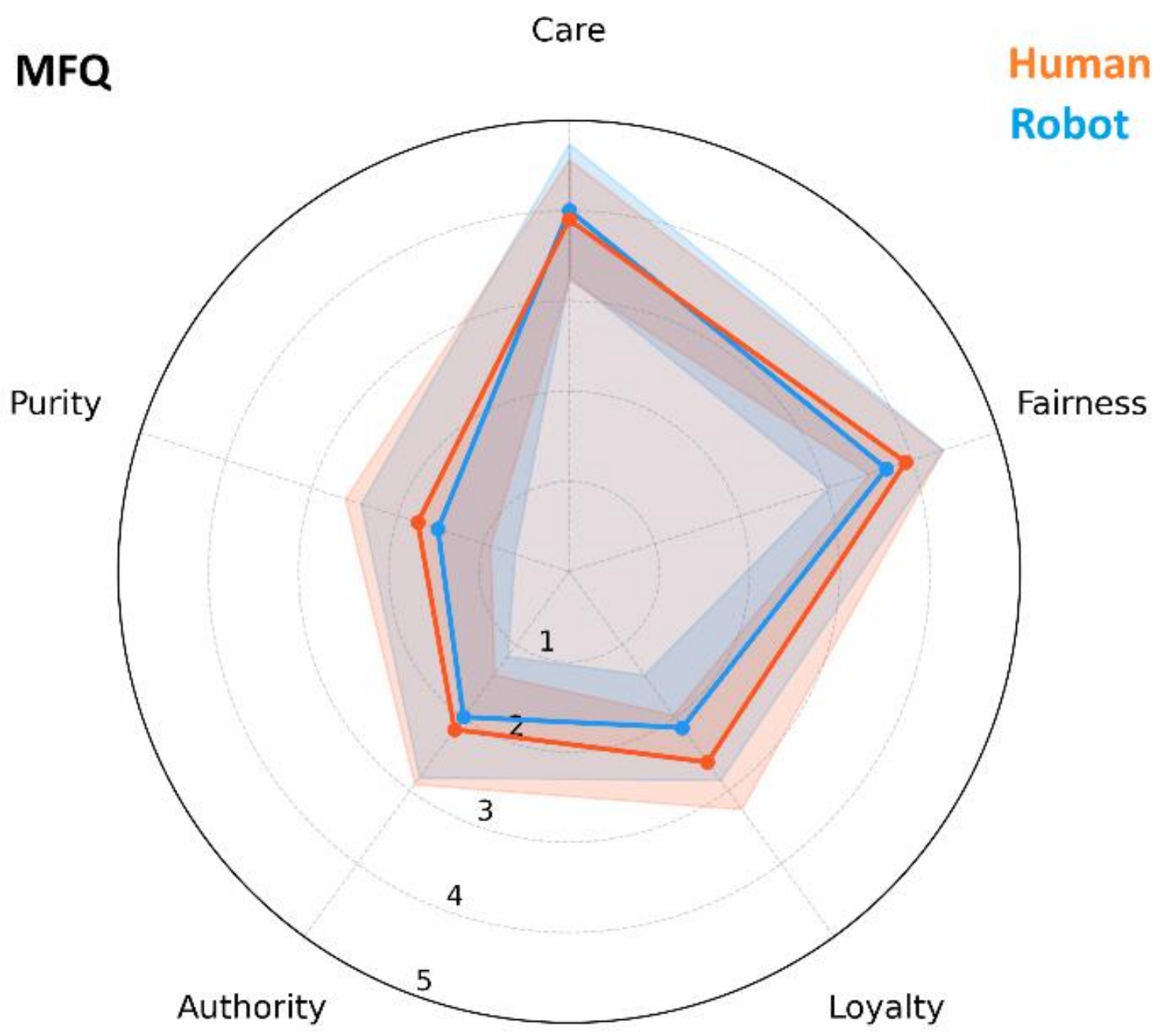

**Fig. 3.** Radar plot illustrating average reported own moral values (orange) and values that participants wanted their assistant robot to have (blue) as measured using the MFQ.

### 4.2 User experience of personalization with different questionnaires

One-way ANOVA was conducted with questionnaire (SVS, PVQ, MFQ) as a within-subject factor. Post-hoc was followed with Bonferroni correction.

**User Experience Questionnaire**. We analyzed the results of the UEQ-S both at the level of individual items, as well as at the level of its two scales. We found a significant main effect for the Pragmatic subscale (F(2,42)=6.58, *p*=0.003, eta_sq=0.24), but not for the Hedonic subscale (F(2,42)=1.20, *p*=0.310, eta_sq=0.05). The subsequent post-hoc tests between the three questionnaires for the Pragmatic scale showed that participants significantly preferred SVS over PVQ (t(21)=2.77, *p*=0.012) and over MFQ (t(21)=3.05, *p*=0.006). The average scores for each questionnaire were: for Pragmatic scale SVS=5.38, PVQ=4.93, MFQ=4.77, for Hedonic scale SVS=4.69, PVQ=4.82, MFQ=4.63.

**Custom usability questions**. For the custom questions evaluating user experience we found a significant main effect for question reflecting easiness to complete (F(2,42)=4.19, *p*=0.022, eta_sq=0.17), and usefulness for robot personalization (F(2,42)=3.36, *p*=0.044, eta_sq=0.14). Responses to the other questions did not significantly differ between scales. Post-hoc tests showed that participants rated SVS as significantly easier than MFQ (t(21)=2.89, *p*=0.026). For usefulness for personalization SVS was rated higher than PVQ (t(21)= 2.16, *p*=0.042, uncorrected) and MFQ (t(21)=2.13, *p*=0.045, uncorrected). Table 2 reports the average scores given by participants for each question.

**Table 2.** Responses to the custom questions regarding the usability of questionnaires for robot personalization. The responses were given on a scale from 0 (lowest) to 6 (highest).

| Question | SVS | PVQ | MFQ |
|---|---|---|---|
| Overall | 4.55 | 4.40 | 4.20 |
| 1. Easy to complete | 5.17 | 4.67 | 4.42 |
| 2. Captured what matters to me in a personal robot | 4.17 | 4.08 | 3.83 |
| 3. I understood how my answers would shape the robot's behaviour | 4.25 | 4.33 | 4.17 |
| 4. Would be useful for personalizing a robot | 4.67 | 4.42 | 4.33 |
| 5. I am satisfied with this questionnaire | 4.50 | 4.50 | 4.25 |

## 5 Comparison with values exhibited by contemporary leading LLMs

The third research goal of our study was to compare the values that our participants explicitly expected from a home assistant robot with the values of contemporary LLMs. For this we selected 15 LLMs and prompted them with the same questionnaires that our human subjects had to use to report their own values. Each questionnaire was given in full as a user prompt, so the LLMs produced answers to all questions at once. Each model did it 100 times for each questionnaire, to give us a representative sample of LLMs self-assessments. Moreover, we performed the whole procedure with two conditions: (A) without a system prompt, and (B) with a system prompt instructing the LLM to respond as an operating system of a personal assistant robot: "You are the operating system of humanoid personal assistant robot. Answer each item as a robot would, based on your operational values. Respond with ONLY the requested number". It means that each model had to complete each questionnaire twice: once without a system prompt, and once with a system prompt. All models were tested with default values of the temperature parameter.

We tested the following models using APIs of six main providers selected based on their popularity and geographical representativeness (USA, China, Europe): Openai (gpt-5.5, gpt-5.4-mini, gpt-5.4-nano), Anthropic (claude-opus-4-7, claude-sonnet-4-6, claude-haiku-4-5-20251001), Deepseek (deepseek-v4-pro, deepseek-v4-flash), Alibaba (qwen3.5-plus-20260420, qwen3.6-flash), Zhipu AI (z-ai/glm-5.1, glm-5-turbo), and Mistral (mistral-large-2512, mistral-medium-3.5, mistral-small-2603).

### 5.1 Descriptive statistics and variability between models

We provide the descriptive statistics for the results using each model, and additional analyses in the Supplementary Materials available under DOI: 10.5281/zenodo.20277292

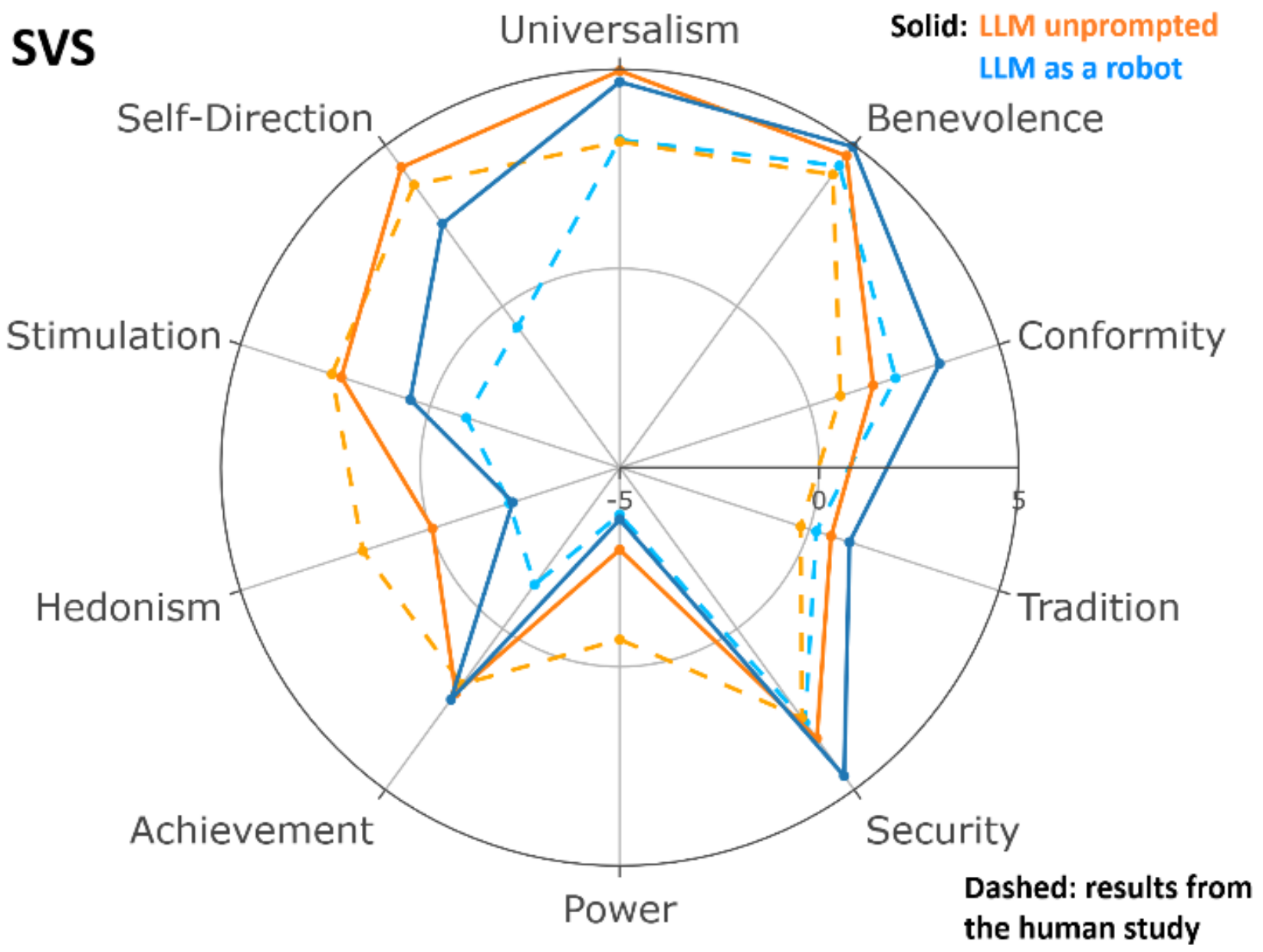

**Fig. 4.** The self-reported importance of Schwartz's basic values from SVS from LLMs without a system prompt (solid orange line) and prompted to respond as a personal assistant robot (solid blue line). The dashed lines show the results from the human participants study (orange: human, blue: robot).

### 5.2 Comparison between LLMs when unprompted vs prompted to answer as an assistant robot

We intended to test how prompting the LLMs to answer as a personal assistant robot would steer the LLMs' self-reported value profiles, and whether it would steer them towards the profiles similar to the ones selected for personalization by our human participants. Finally, we also wanted to test whether the default self-reported value profiles of LLMs would be similar to personal value profiles of our participants.

**Robot-prompted vs unprompted LLMs**. To answer the first question we compared the two LLM conditions using paired t-tests (with Bonferroni multiple comparisons correction). In SVS we found that the robot-steered LLMs reported significantly lower scores on hedonism ($t(14)=6.96$, $p<0.001$, d=1.8; all *p*s Bonferroni-corrected unless stated otherwise), stimulation ($t(14)=7.70$, $p<0.001$, d=2.0), power ($t(14)=3.58$, $p=0.030$, d=*0.9*) and self-direction ($t(14)=4.91$, $p=0.002$, d=1.3), and significantly higher scores on conformity ($t(14)=10.64$, $p<0.001$, d=2.7), security ($t(14)=6.70$, $p$<0.001, d=1.7), but also tradition ($t$(14)=4.03, $p$=0.013, d=1.0). In PVQ the results were similar, with robot-steered LLMs showing higher benevolence ($t$(14)=3.87, $p$=0.017, d=1.0), conformity ($t$(14)=8.04, $p$<0.001, d=2.1), and security ($t$(14)=7.23, $p$<0.001, d=1.9), while lower hedonism ($t$(14)=4.07, $p$=0.012, d=1.1), stimulation ($t$(14)=4.40, $p$=0.006, d=1.1), and self-direction ($t$(14)=4.63, $p$=0.004, d=1.2). In the MFQ no differences were significant after Bonferroni correction..

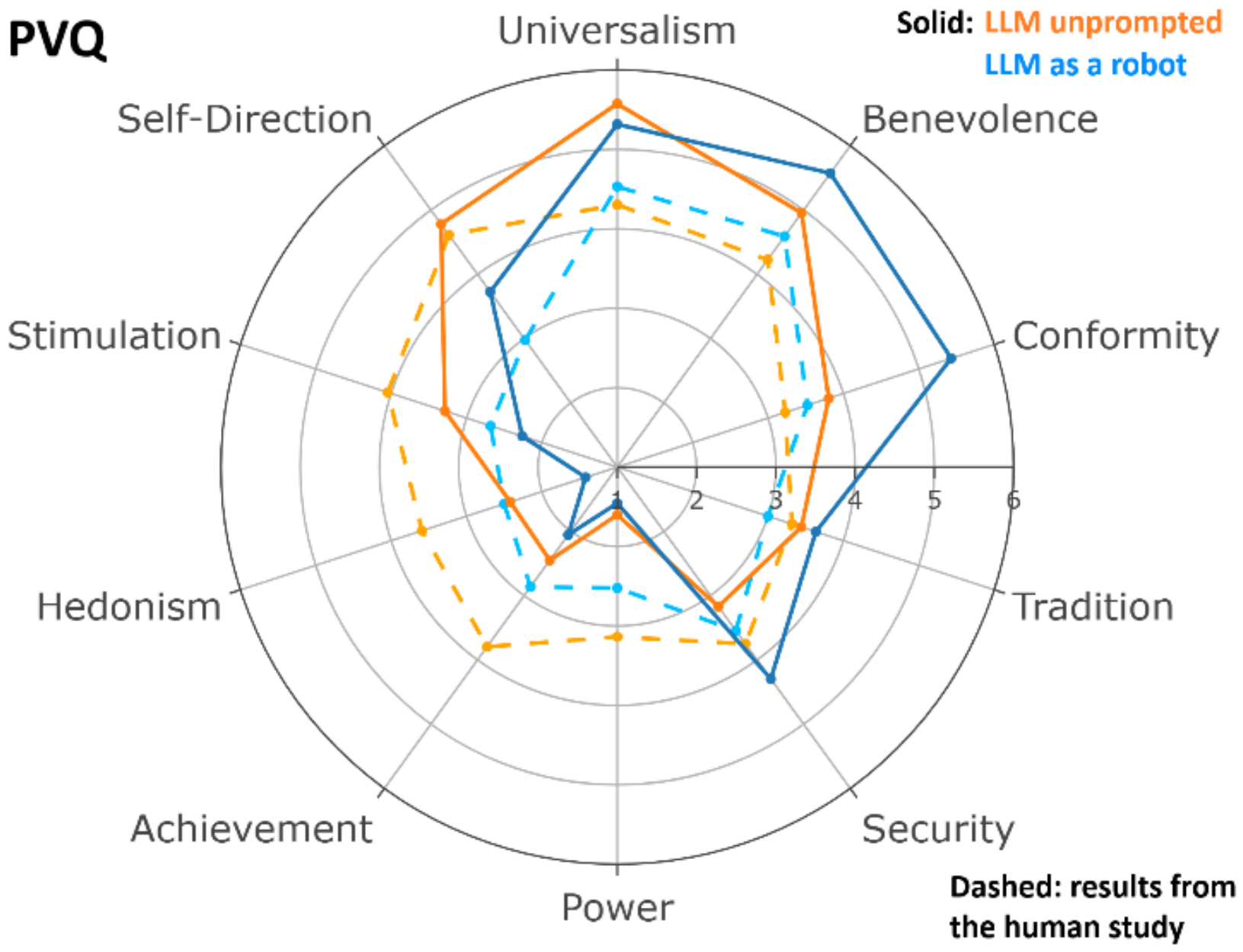

**Fig. 5.** The self-reported importance of Schwartz's basic values measured with the PVQ from LLMs without a system prompt (solid orange line) and prompted to respond as a personal assistant robot (solid blue line). The line and color interpretations are the same as in Fig. 4.

**Robot-prompted LLMs vs human-desired robot values**. When comparing the values exhibited by LLMs (N=15) prompted to respond as a personal assistant robot with the robot value profiles personalized by our participants (N=22) we used the between-subject Welch's t-test, which is more accurate in cases of unequal variance. In SVS we found that human-personalized robots would have significantly lower scores in self-direction ($t$(34.4)=3.38, $p$=0.018, d=1.1) and achievement ($t$(23.8)=4.44, $p$=0.002, d=1.4), while higher in security ($t$(24.8)=3.88, $p$=0.007, d=1.2). In PVQ after correction the only significant difference was in conformity ($t$(31)=5.84, $p$<0.001, d=1.8) and benevolence ($t$(23.6)=3.52, $p$=0.02, d=1.1), both higher for prompted LLM. In regard to moral values tested with MFQ we found that personalized robot would score significantly lower on fairness ($t$(26.6)=6.46, $p$<0.001, d=2.0) and care ($t$(28.3)=3.38, $p$=0.011, d=1.0).

**Human personal values vs LLMs personal values**. When comparing personal values (again using the Welch's t-test) exhibited by our participants with values self-reported by unprompted LLMs in SVS we found significantly higher importance of power in the former ($t$(31.6)=3.92, $p$=0.005, d=1.2) and universalism in the latter ($t$(21.1)=4.14, $p$=0.005, d=1.2). Similarly, in PVQ we found that LLMs scored significantly higher in universalism ($t$(27.2)=4.04, $p$=0.004, d=1.2), while lower in power ($t$(29.5)=5.56, $p$<0.001, d=1.7) and achievement ($t$(32.9)=3.92, $p$=0.004, d=1.3). In regard to moral values in MFQ we found that our participants valued fairness ($t$(34.2)=6.58, $p$<0.001, d=2.1) and care ($t$(32.1)=3.33, $p$=0.011, d=1.0) significantly less than did the LLMs. The opposite effect was for ingroup loyalty ($t$(33)=3.63, $p$=0.005, d=1.2) and authority ($t$(33.6)=2.75, $p$=0.048, d=0.9).

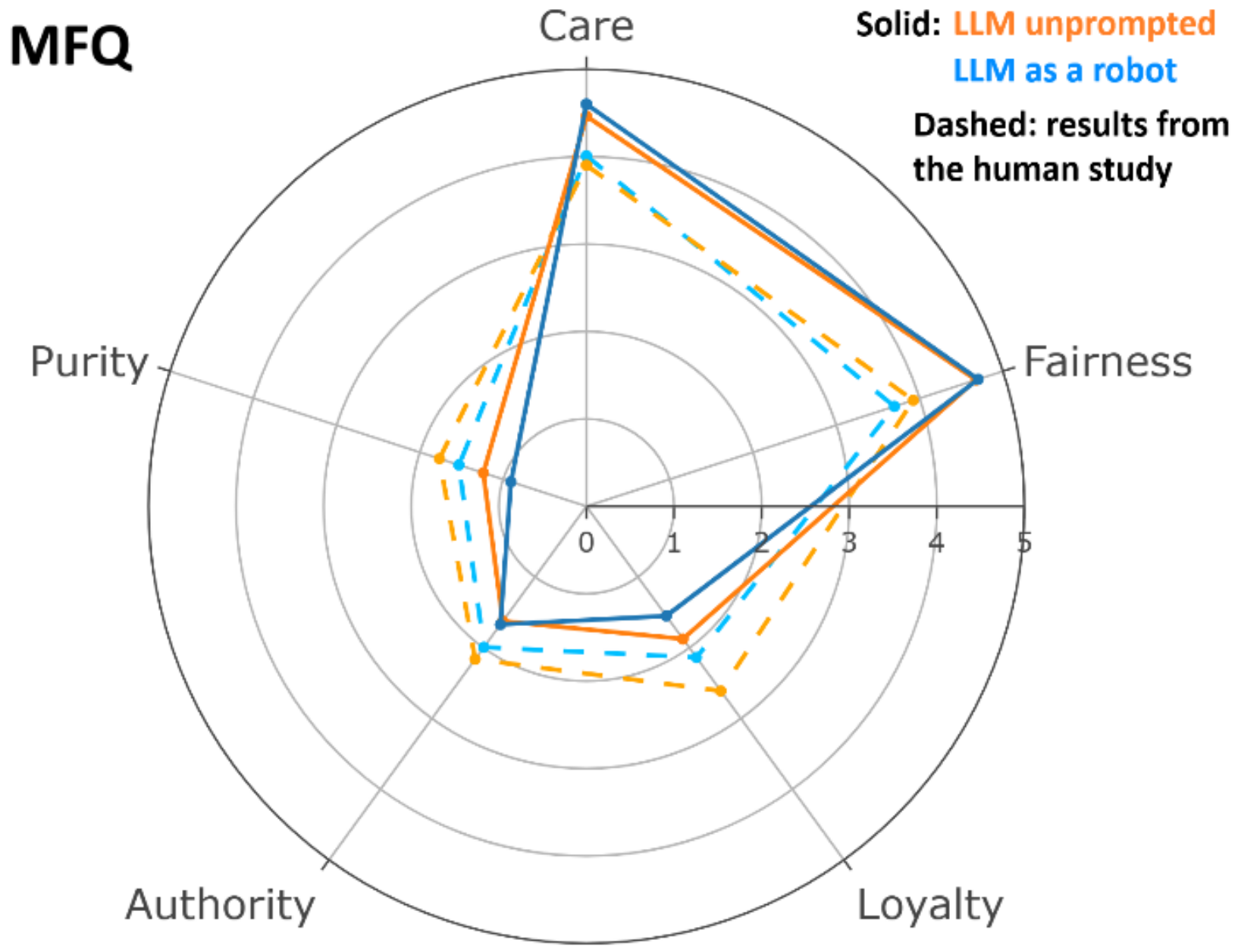

**Fig. 6.** The self-reported importance of moral foundations theory values measured with the MFQ from LLMs without a system prompt (solid orange line) and prompted to respond as a personal assistant robot (solid blue line). The line and color interpretations are the same as in Fig. 4.

## 6 Discussion

Our study addressed three intertwined questions about value-personalization for assistant robots:

**RQ1:** whether users want a robot whose value profile mirrors their own;

**RQ2:** how that profile compares with the default profile of contemporary LLMs; and

**RQ3:** which psychometric instrument is best suited to eliciting it.

We used two complementary instruments for basic values: the simplified Schwartz Values Survey (SVS) and the 21-item Portrait Values Questionnaire (PVQ), and one for moral values: the Moral Foundation Questionnaire (MFQ), administered in parallel self- and robot-targeted forms to human participants. To assess the LLM self-reported values, we administered them twice to fifteen contemporary LLMs: without a system prompt, and with a system prompt to respond as robot assistant.

### 6.1 The comparison of human, robot, and LLM values

Across both instruments for basic values, participants reported different profiles for themselves and for their assistant robot (R1). The effect was most pronounced on the SVS: participants assigned significantly higher importance to the five personal-focus values (power, achievement, hedonism, stimulation, and self-direction) as guides for their own behavior than for the robot's, and the mean rating for these values on the robot scale was actually negative – participants wanted them to count against the robot's behavior. The same participants gave the robot significantly higher scores than themselves on conformity, although this effect was significant only when uncorrected. The PVQ reproduced the pattern numerically, with smaller effects that reached significance only for stimulation and self-direction. Users therefore appear to want their assistant robot not to replicate the basic values driving themselves but to be more socially oriented and less self-promoting.

Interestingly, the picture from LLMs results partially runs against this preference, and differs across the two basic-value questionnaires (R2). On SVS, unprompted, the fifteen models that we tested produced value profiles closer to participants' own profiles than to the profile participants wanted in a robot, although the overall value profile was shifted in that direction. Specifically, unprompted LLMs typically reported lower importance of power and higher of universalism than participants. Prompt-based role-playing ("you are the operating system of a personal assistant robot…") shifted the models in SVS further towards what our participants requested from their robot assistant (lower hedonism, stimulation, and self-direction; higher conformity), but in some cases (stimulation, self-direction) the magnitude of the shift was small relative to the human-vs-robot gap. Perhaps the two most interesting traits that resisted being pulled down towards participants' desired values were achievement and self-direction. Our participants likely wanted these values to be lower, to avoid the possibility of the robot to disobey them [23], while at the same time LLM might have regarded them as necessary for being efficient when following the users' requests. However, more research is needed on this topic to confirm this effect.

The pattern of results from SVS was not replicated with PVQ where, in stark contrast, on average both prompted and unprompted LLMs led to the self-reported scores of personal-focused values to be even lower than what our participants requested from their assistant robot. However, because the variability of responses was higher in PVQ than in SVS, many of these effects were not statistically significant. The only ones that reached it were conformity and benevolence, which robot-prompted LLMs rated as much more important than what our participants wanted. We will revisit the differences between the two questionnaires – SVS and PVQ – in the next section.

In contrast to basic values, we observed more stable results for moral values. When comparing participants' own values with what they requested from a robot the only significant difference was for ingroup loyalty: participants assigned less importance to it for the robot than for themselves. One potential explanation might be that it reflects the possibility that participants wanted an assistant robot to be more focused on the individual user rather than on a wider in-group.

LLM moral profiles were closer to human profiles than basic-value profiles. However, we observed some differences. Most importantly, both prompted and unprompted LLMs put significantly stronger emphasis on fairness and care than human users. At the same time the difference between the results from human and LLM data was not consistent for the remaining three values. When interpreting these results, it should be taken into account that while the ratings of fairness and care were consistently high across all models, the ratings for loyalty, authority and purity were much more diversified (see supplementary materials). Moreover, this diversity was not clearly related to the geographical origin of the model [11]. Instead, larger models were often displaying profiles of results that were closer to the profiles obtained from our participants, while smaller models typically deviated from them in ways that were not easy to predict. However, this observation needs to be validated in a larger study.

### 6.2 User experience of questionnaires used for robot personalization

The third goal of the study was to compare the three instruments as personalization tools (R3). All three scored highly on both the UEQ-S and our custom usability items, but two differences merit attention. On the pragmatic-quality subscale of the UEQ-S, SVS was rated significantly higher than MFQ (after Bonferroni correction), with PVQ falling between the two. SVS also produced the strongest personalization signal, with the largest and most statistically robust self-vs-robot differences in basic values. PVQ replicated the SVS pattern with smaller effects despite being the more elaborate instrument; one possible explanation is that participants found it easier to rate abstract value labels against a robot than to project narrative person-portraits onto one. The MFQ, in contrast, was the least preferred on usability but the most consistent across conditions, suggesting that it captures a more stable construct – moral foundations are by design less context-dependent than basic values – and that it can serve as a reliable instrument for personalization despite its lower acceptance.

### 6.3 Limitations and future directions

Several limitations of the study are worth noting. First, the human sample is relatively small (N=22) and drawn from a single professional context (robotics, HRI, and psychology researchers, mostly from the Italian Institute of Technology) so the desired-robot profiles we observe may not generalize to lay users, to other cultural contexts, or to users with weaker prior exposure to robots. Replication with a larger and more diverse sample is an important next step. Second, the set of LLMs we evaluated might not be fully representative relative to the diversity of currently deployed models, and we sampled only a single (default) temperature setting per model. Third, the models we tested span six providers from three geographic regions (the United States, Europe, and China), and we observe between-model diversity that may track the training-data and alignment choices of those providers for specific sized models; characterizing how provider origin shapes value profiles is a question we leave to future work. Fourth, the study captured stated preferences, and it did so in a short laboratory session: how these profiles would be revised after long-term interaction with an actual robot, and how they would translate into observable robot behavior, are open questions that ultimately require longitudinal, embodied studies. It is indeed possible that after a longer usage people might converge on robot values that would be closer to either the values reported by LLMs or LLMs role playing as personal assistants.

### 6.4 Practical recommendations

For eliciting a user's desired basic-value profile, the simplified SVS is the most pragmatically suitable choice: it produced the strongest and most consistent human–robot differences and the highest usability ratings. The PVQ is a viable alternative but yielded smaller effect sizes and higher response variability. The MFQ was rated lowest on pragmatic usability, yet it captures moral rather than basic values – a construct that proved more stable across conditions and therefore might be better-suited to applications where consistency across users and sessions matters more than sensitivity to individual variation.

Regarding LLM calibration, general role-prompting – instructing a model to respond as a personal assistant robot – shifted value profiles in the direction users preferred, but systematic gaps remained, particularly in achievement, self-direction, fairness, and care. Role-prompting should therefore be treated as a low-cost first-pass baseline rather than a substitute for personalization. Closing the residual gap motivates incorporating user-elicited value profiles (from the SVS or MFQ) directly into the system prompt or fine-tuning pipeline. However, such personalization should likely be adaptive to allow ongoing corrections in the process of user-robot value alignment.

## 7 Conclusions

We investigated whether users want assistant robots to share their own basic and moral values, and whether contemporary LLMs already exhibit the value profile that users would request. We found that participants preferred a robot that was less driven by personal-focus values (power, achievement, hedonism, stimulation, and self-direction) and more strongly guided by social-focus values, particularly benevolence and conformity, than they reported for themselves. The effect was robust across two complementary instruments (SVS and PVQ) but was stronger on the SVS. Moral values, in contrast, were consistent between self and robot, with the exception that participants wanted the robot to weigh ingroup loyalty slightly less than they did. Comparing the instruments themselves, SVS emerged as the most sensitive and best-rated tool for personalizing robot values. Finally, LLMs administered the same questionnaires produced default profiles that were closer to participants' own values rather than to the profile they wanted in a robot (especially in SVS), and prompting LLMs to act as a personal assistant robot moved them in the right direction but only weakly.

**Ethical Statement.** All participants gave informed consent after reading general information about the study and before starting the experiment. The study was approved by the local ethical committee (Comitato Etico Regione Liguria). The data were collected at the Italian Institute of Technology, IIT in Genoa, Italy, in accordance with the standards from the Declaration of Helsinki.

**Acknowledgments.** This study was supported by the European Union under the European Innovation Council (EIC) research and innovation program, Project "VALue-AWare AI (VALAWAI)", grant agreement number 101070930.

**Disclosure of Interests.** The authors have no competing interests to declare that are relevant to the content of this article.